\documentclass[aps, prl, amsmath, floats,floatfix, twocolumn, superscriptaddress]{revtex4}

 \usepackage{xcolor}
\usepackage{amssymb}
\usepackage{amsmath}
\usepackage{verbatim}
\usepackage{mathrsfs}
\usepackage{amsfonts}
\usepackage{latexsym}
\usepackage{epsfig}
\usepackage{color}
\usepackage{graphicx,subfigure}
\usepackage{units}
\usepackage[spanish,english]{babel}

\usepackage{mathtools,slashed}

\newcommand{\be}{\begin{equation}}
\newcommand{\ee}{\end{equation}}
\newcommand{\bea}{\begin{eqnarray}}
\newcommand{\eea}{\end{eqnarray}}

\begin{document}

\title{Predictions for the CMB from an Anisotropic  Quantum  Bounce}

\author{Ivan Agullo}
\email{agullo@lsu.edu}
\author{Javier Olmedo}
\email{jolmedo1@lsu.edu}
\affiliation{Department of Physics and Astronomy, Louisiana State University, Baton Rouge, LA 70803, U.S.A.
}
\author{V.~Sreenath}
\email{sreenath@nitk.edu.in}
\affiliation{Department of Physics, National Institute of Technology Karnataka, Surathkal, Mangalore 575025, India.}


\begin{abstract}

We introduce an extension of the standard inflationary paradigm on which the big bang singularity is replaced by an anisotropic  bounce.  
Unlike in the big bang model, cosmological perturbations find an adiabatic regime in the past. We show that this scenario accounts for the observed quadrupolar modulation in the temperature anisotropies of the cosmic microwave background, and we make predictions for the polarization angular correlation functions $E$-$E$, $B$-$B$ and $E$-$B$, together with temperature-polarization correlations $T$-$B$  and $T$-$E$, that can be used to test our ideas. We base our calculations on the bounce predicted by loop quantum cosmology, but our techniques and conclusions apply to other bouncing models as well. 

\end{abstract}

\maketitle

{\bf Introduction.} Anisotropies are generic features of homogeneous solutions to Einstein's equations. This is manifest already for Bianchi I geometries, the simplest anisotropic spacetimes. There, in the absence of anisotropic sources, the contribution of shears to Friedmann equations dilutes with the expansion faster than that of matter and radiation. Therefore, unless anisotropies are exactly zero during the entire history of the cosmos, there must be a time in the past when they were dominant. 
From this viewpoint, the Friedmann-Lema\^itre-Robertson-Walker (FLRW) isotropic spacetimes are quite singular. 
 In the standard model of cosmology, one appeals to a phase of slow-roll  inflation, when the  
  exponential expansion quickly  dilutes anisotropies, and argues that from that time on one can just ignore them. However, the way this argument is applied contains a stronger assumption---that the quantum states describing cosmological perturbations  were also isotropic. 
Anisotropies in quantum fields do not dilute at the same rate as the shears of the homogeneous metric do. In fact, the only reason why they can be washed away is because the cosmic expansion redshifts the wavelengths for which the perturbation fields are anisotropic, potentially shifting them out of the observable Universe. There is no additional dilution \cite{agulloparker}. However, redshift scales linearly with the expansion, while the dilution of the shear $\sigma^2$ scales with its sixth power (in absence of anisotropic sources). Hence, unless inflation is significantly longer than the minimum amount required, one cannot rule out that some anisotropic features were imprinted in the cosmic microwave background (CMB). 

This argument, and the fact that the {\it Planck} satellite has observed anisotropies in the  CMB \cite{plnck2018}, has triggered our interest in studying anisotropic extensions of the standard cosmological model. However, within general relativity one finds a major impediment: In a generic anisotropic universe, there are no preferred initial states for the cosmological perturbations. In the theory of inflation, one uses the fact that the wavelengths of the perturbations that we can probe in the CMB were much shorter than the Hubble radius at the onset of slow roll. Then, the notion of adiabatic vacuum can be used to single out an initial quantum state, at least for these wavelengths. However, this argument fails if the preinflationary spacetime is anisotropic (see e.g.\ \cite{ppu-BI1}). In the absence of preferred initial data, the theory loses predictive power.

This Letter proposes an extension of the standard model, where the big bang singularity is replaced by an {\it anisotropic cosmic bounce}. We consider a framework in which the Universe contracts in the remote past, according to Einstein's theory, until matter and spacetime curvature  approach the Planck scale. Then, quantum gravity effects grow and  dominate the dynamics, overwhelming the classical attraction and making the Universe  bounce.  In the far past, the Universe isotropizes and perturbations find an adiabatic regime. 
Therefore, in this scenario one has preferred initial and final notions of vacua and Hilbert spaces for perturbations. Our goal is to formulate  this quantum  theory and to solve the evolution, that in the Schr\"odinger picture reduces to compute the $\mathcal{S}$ matrix  between  $in$ and $out$ states.  We show that  perturbations can retain memory of the anisotropic phase of the Universe, and leave an imprint on the CMB, even though anisotropies in the background  metric are large only during a short period of time around the bounce. In order to isolate the effects of anisotropies, we work with Bianchi I spacetimes; they differ from spatially flat FLRW spacetimes only by the presence of anisotropic shears.

{\bf The classical phase space.} Loop quantum cosmology (LQC) uses canonical methods for quantization \cite{asrev,lqc}. Therefore, to incorporate perturbations we first need to formulate them in the Hamiltonian language. 
This task is significantly more tedious and complex than the FLRW counterpart \cite{lang}, and to the best of our knowledge it has not been developed before (although classical gauge invariant perturbations in Bianchi I have been studied in \cite{ppu-BI1} by expanding Einstein's equations).  We follow the geometric approach proposed in \cite{gnr}. Gauge invariant fields at linear order in perturbations can be obtained by finding a canonical transformation that makes four of the new momenta proportional to each of the four linear constraints of the theory, respectively---the scalar and vector constraints. This  guarantees that the conjugate variables to these momenta are pure gauge, while the rest of fields are gauge invariant. The search for such  transformation reduces to solving Hamilton-Jacobi-like equations for a generating function.  There are multiple solutions, which correspond to different choices of gauge invariant fields. We have selected the choice that in the isotropic limit  reduces to the familiar scalar perturbations and the two circularly polarized tensor modes (with helicity $\pm2$), and denote them by $\Gamma_0$ and $\Gamma_{\pm 2}$, respectively. 

The dynamics of gauge invariant perturbations is guaranteed to decouple from pure gauge fields, and is generated by a Hamiltonian $\mathcal{H_{\rm pert}}$. Hamilton's equations can be combined into the second-order differential equations 
\be  \label{eqginper}\ddot \Gamma_{s}+3\, H\, \dot \Gamma_{s}+\frac{k^2}{a^2}\,  \Gamma_{s}+\frac{1}{a^2}\, \sum_{s'=0}^2\, {\cal U}_{ss'}\, \Gamma_{s'}=0\, , \ee
with $ s=0,\pm2$;  we have expanded the fields in Fourier modes  $\Gamma_{s}(\vec k,t)$, and $k$ is the comoving wavenumber. The functions ${\cal U}_{ss'}(\vec k,t)$ are effective potentials made of  a complicated combination of the background variables (see \cite{aos} for details), $a(t)$ is the mean scale factor, and $H=\dot a/a$ its Hubble rate. We have implemented this Hamiltonian theory in the symbolic language  of {\tt Mathematica}, and made the code publicly available in \cite{math-nb}. One important difference with FLRW spacetimes is that the potentials ${\cal U}_{s,s'}$ are not diagonal in presence of anisotropies. Therefore, the three  fields $ \Gamma_{s}$ are coupled and, because these couplings are time dependent,  there is no way to diagonalize the equations of motion at all times by means of a local field redefinition.

{\bf Quantum theory.} 
The classical phase space we are interested in is the product ${\bf V_{\rm BI}}\times{ \bf V_{\rm pert}}$ of  Bianchi I geometries and gauge invariant perturbations. At  leading order in the perturbations, dynamics is implemented by first determining the evolution within ${\bf V_{\rm BI}}$, and then lifting the dynamical curves to ${ \bf V_{\rm pert}}$ with the Hamiltonian $\mathcal{H_{\rm pert}}$. We follow the same strategy in the quantum theory. Namely, the Hilbert space is the product ${\bf H_{\rm BI}}\otimes{ \bf H_{\rm pert}}$.  ${\bf H_{\rm BI}}$ has been described in \cite{awe,mmwe}. A good approximation for quantum states $\Psi_{\rm BI}\in {\bf H_{\rm BI}}$ that at late times are sharply peaked on a classical  geometry is provided by the so-called effective equations \cite{chiou-vandersloot}. These are quantum corrected equations for the directional scale factors and their conjugate variables, whose solutions follow with precision the peak of the wave function  $\Psi_{\rm BI}$. The physics of these spacetimes has been studied in detail in \cite{gs}, and the main features are the following. 
All solutions  contain  a bounce of the mean scale factor $a(t)$,  which is  caused by quantum gravity effects.  All strong curvature singularities are resolved, as long as the matter sector satisfies the null energy conditions. Energy densities and shears  are bounded from above. Directional scale factors $a_i(t)$  bounce generically at different times, giving rise to  a richer bounce than in the isotropic case.  After the bounce, and in the presence of a scalar field and  an inflationary potential $V(\phi)$,  Hubble friction slows $\phi$ down  and generically leads to a phase of slow roll; such  a phase is  an attractor in the phase space of this quantum corrected  theory \cite{gs}. In this sense, the bounce provides a mechanism to set up the initial conditions for inflation to occur. Once inflation starts, the scenario provided by the standard cosmological model goes through, with the important difference that the state of perturbations is different from the standard ansatz of the Bunch-Davies vacuum.

We assume the matter content to be a scalar field with a potential $V(\phi)$.  In the scenarios of interest  (see below) the potential is subdominant in the preinflationary phase, and consequently the generation of anisotropies is independent of $V(\phi)$. For the sake of simplicity, we use $V(\phi)=1/2\, m^2 \phi^2$ and comment below on the  effect of other choices.  The other freedoms in our predictions come from the choice of an effective Bianchi I quantum spacetime. One such geometry is  singled out by specifying the value of the shear squared $\sigma^2(t_B)$, the shear in one of the principal directions, say $\sigma_x(t_B)$,  the value of the scalar field $\phi(t_B)$, and the sign of its time derivative,  all at the time $t_B$ of the bounce. $\sigma^2(t_B)$ measures the  total amount of anisotropies at $t_B$;  $\sigma_x(t_B)$ indicates the way these anisotropies are distributed in the three principal directions, and $\phi(t_B)$ and the sign of $\dot \phi(t_B)$ control the number $N$ of $e$-folds of expansion from the bounce to the end of inflation [$\sigma^2(t_B)$ also affects this number, but in a subleading manner \cite{gs}.

To quantize the perturbations, we follow the conceptual framework introduced in \cite{akl,aan,hyb} and extend it to  Bianchi I geometries. We obtain that the dynamics of quantum perturbations  $\hat \Gamma_0$, $\hat \Gamma_{\pm 2}$  are described by the  equations (\ref{eqginper}), with the background geometry given by a solution to the effective equations of LQC.  The main difficulty arises from the  interactions among the quantum fields $\hat \Gamma_0$, $\hat \Gamma_{\pm 2}$, induced by the anisotropies.  To describe dynamics, we first define the $in$ and $out$ Hilbert spaces. The former is defined from an adiabatic vacuum in the past (see Sec.\ IV in \cite{aos} and \cite{aos2} for details), that we take to be anytime before $10000$ Planck times prior to the bounce. At this time, anisotropies are already negligible in the geometries that we have explored, and all Fourier modes of interest  are well inside the Hubble radius. The $out$ Fock space is the standard one built from the Bunch-Davies vacuum during inflation, when  the anisotropies of the spacetime are negligible again. The quantum evolution is implemented by the $\mathcal{S}$ matrix, which provides a unitary map between the $in$ and $out$ Fock spaces \cite{aa}. Its action on the $in$ vacuum produces
 \begin{align}  \label{Sinout2} & \hat{\mathcal{S}}|in\rangle =\\  &\bar N\, \bigotimes_{\vec k} \exp{\Big[\sum_{s,s'=0,\pm2}  {\rm V}_{ss'}(\vec k) \, \hat  a^{out\, \dagger}_{s}(\vec k)\  \hat a^{out \, \dagger}_{s'}(-\vec k)\Big]} \, |out \rangle \nonumber  \, , \end{align} 
where $\bar N$ is a normalization factor, and $V_{ss'}(\vec k):=\sum_{s''} \frac{1}{2} \, \beta^{*}_{s'' s}\, (\alpha^{-1})^*_{s's''}$, with $\alpha_{ss'}(\vec k)$ and  $\beta_{ss'}(\vec k)$ the Bogoliubov coefficients that relate the $in$ and $out$ vacua. They encode the information of the evolution of perturbations across the anisotropic bounce, and  can be computed from the classical equations of motion. 
The operators $\hat  a^{out\, \dagger}_{s}$,  with $s=0,\pm2$, create quanta of the familiar scalar and tensor modes in inflation, respectively. The right-hand side of (\ref{Sinout2}) is the product of squeezing operators acting on $|out\rangle$. Consequently, the $in$ vacuum evolves to a state made of entangled  pairs of quanta, one with  wave number $\vec k$ and the other with  $(-\vec k)$; i.e., no net momentum is created. In the isotropic limit $V_{ss'}$ becomes diagonal, and  the operator in (\ref{Sinout2}) becomes the product of  operators for scalar and each of the two tensor modes. This is not the case in presence of anisotropies, where the final state contains entanglement among the three types of perturbations. One can compute, e.g., the entanglement entropy, from the Bogoliubov coefficients \cite{aos}.

\begin{figure}[h]
{\centering     
\includegraphics[width = 0.5\textwidth]{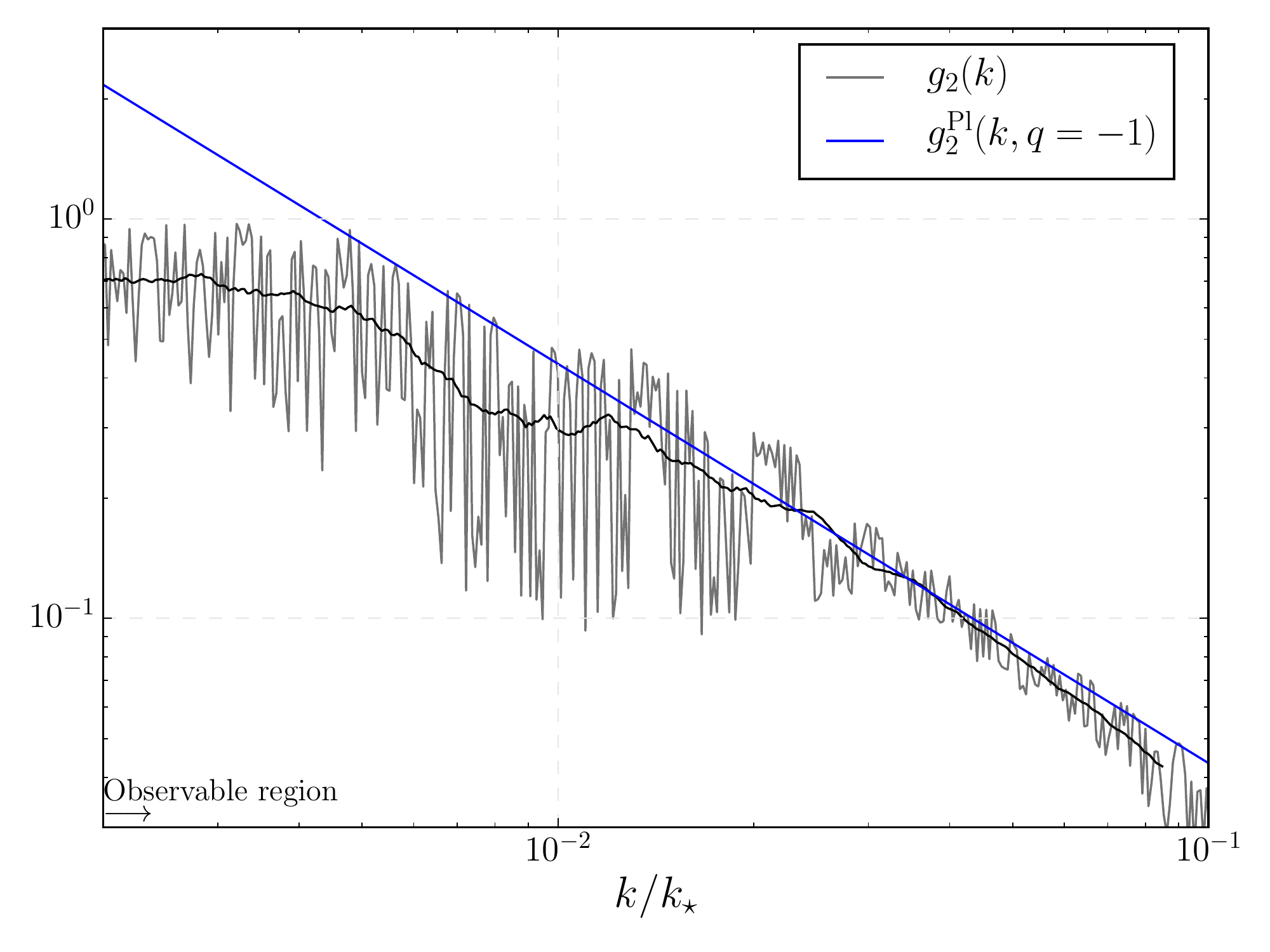}  
}
\caption{Amplitude of the quadrupolar modulation $g_2(k)\equiv \sqrt{\sum_M |g_{2M}|^2/5}$ of the primordial scalar power spectrum $ {\mathcal P}(\vec k) $,  where $g_{2M}(k)=\frac{1}{\bar{\mathcal{P}}(k)} \int d\Omega_{\vec k} \, {\mathcal P}(\vec k)\,  Y^{*}_{2 M}(\hat k)$, with $\bar{\mathcal{P}}(k)\equiv  \int d\Omega_{\vec k} \, {\mathcal P}(\vec k)$. {\it Planck}'s results \cite{plnck2018} for the amplitude of a quadrupole that falls off with $k$ as $g^{\rm Pl}_2(k)= g^{\rm Pl}_2\times (k/k_{\star})^{q}$, for $q=-1$, is shown in  blue. The gray  line shows our results for a set of individual values of $k$. The outcome oscillates with high frequency around the mean value, shown in black.  These oscillations do not show up in angular correlation functions, since they get effectively averaged out when integrating in $k$.  $k_{\star}$ is a reference wave number, whose physical value today is $0.05\, {\rm Mpc}^{-1}$. This plot  is obtained for $\sigma^2(t_B)=5.78$ and $\sigma_{x}=0$ in natural units, and $N=70.1$.} 
\label{fig:g22M}
\end{figure}

{\bf Constraints from observations.}
We next  analyze  {\it Planck}'s observations   of  a quadrupolar direction-dependent modulation in the CMB \cite{plnck2018}.  Since our goal is to describe the largest possible signal that we can expect in the CMB, we choose $\sigma^2(t_B)$  close to its upper bound, and derive the constraints from observations  on the other parameters that specify the spacetime geometry.  Observations translate to a lower bound for the number of $e$-folds $N$, which keeps anisotropies in the CMB below the observed threshold. On the other hand, if this number happens to be very large,  all anisotropies in perturbations would be red-shifted out of the observable Universe. A representative example of our analysis is obtained by choosing $\sigma^2(t_B)=5.78$ in natural units (this is half of its upper bound \cite{gs}) and  $\sigma_x(t_B)=0$. We have computed the quadrupolar modulation and compared it with data from {\it Planck}  (see  Fig.\ \ref {fig:g22M}). The result of this analysis is a lower bound for $N$ of $70.1$. Interestingly, this value is compatible with the results found in \cite{barrau} for  the preferred value of $N$ in anisotropic LQC. As we will shortly see, $N=70.1$ is {\em not} large enough to wash away all anisotropies in the CMB.

{\bf Predictions for the CMB.} We compute the angular correlation functions $C_{\ell \ell^{\prime},m m^{\prime}}^{X, X^{\prime}}\equiv \left\langle a_{\ell m}^{X} a_{\ell^{\prime} m^{\prime}}^{X^{\prime} }\right\rangle$, with 
\be a^X_{\ell m}=\int d\Omega\,  X(\hat n)\,  Y^*_{\ell m}(\hat n)\, ,\ee
where $X=T,E,B$ represents the temperature, electric and magnetic components of the polarization, respectively, of the anisotropies in the CMB.
 
\begin{figure}[h]
{\centering     
\includegraphics[width = 0.5\textwidth]{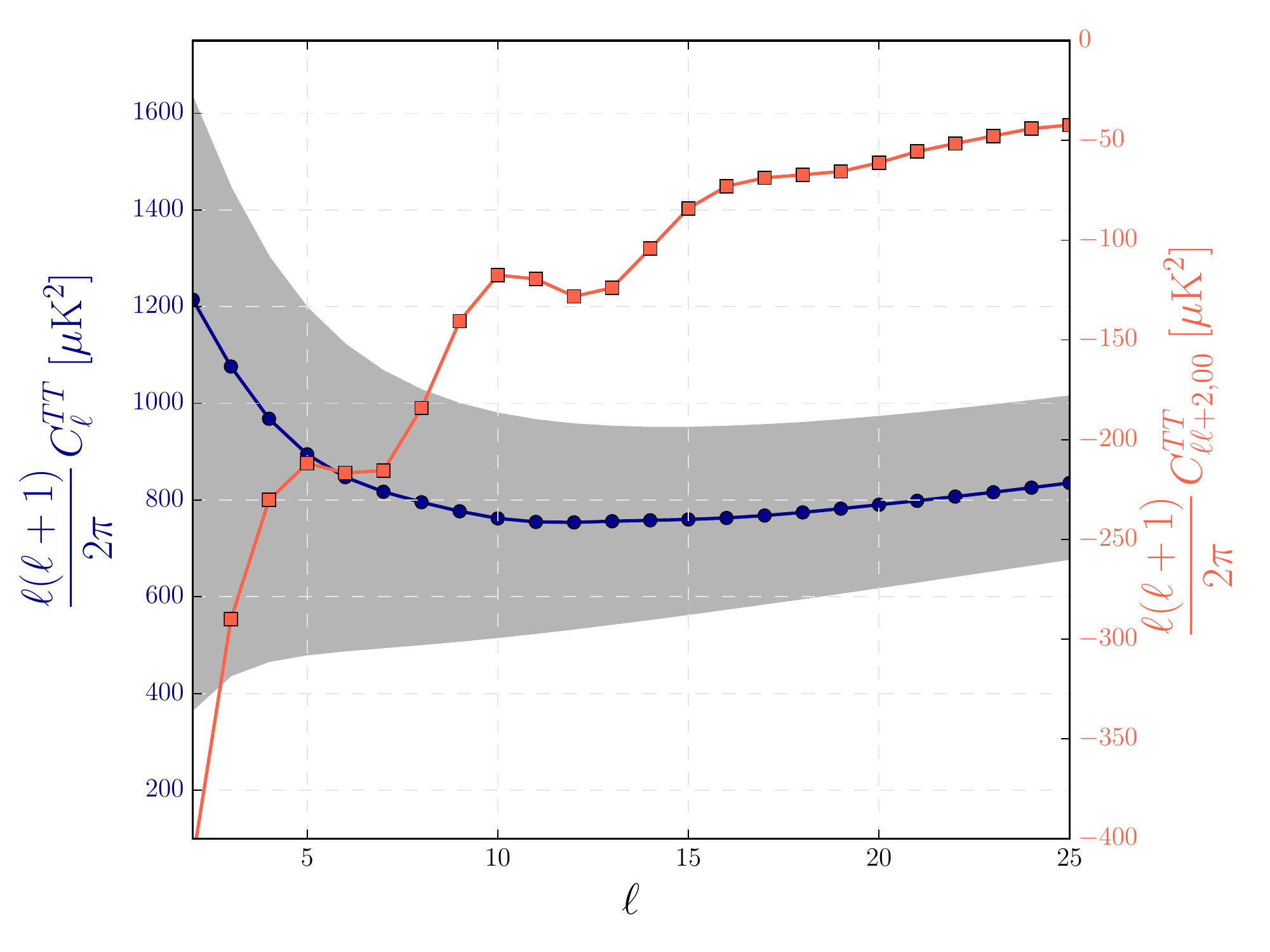}  
}
\caption{Left axis: Temperature-temperature angular correlation function $C_{\ell}^{TT}$ (dotted blue line). For comparison, the shaded region  shows the values obtained from isotropic inflation, including the uncertainties originated from cosmic variance. 
Right axis:  $C_{\ell \ell+2, 00}^{TT}$  (red line with squares) (the isotropic counterpart is exactly  zero).}
\label{fig:DTT}
\end{figure}

(i) Temperature-Temperature ($T$-$T$). Our theory is invariant under translations and parity, but not under rotations. Parity invariance restricts $C_{\ell \ell^{\prime},m m^{\prime}}^{T, T}$ to vanish unless $\ell+\ell'$ is even (isotropy would have also imposed $\ell=\ell'$, $m=-m'$).  We plot in Fig.~\ref{fig:DTT} $C_{\ell}^{TT}\equiv \frac{1}{2\ell+1} \sum_{m=-\ell}^\ell (-1)^mC_{\ell \ell, m -m}^{TT}$, and compare it with the predictions of isotropic  inflation. As expected, the effects of the preinflationary physics are larger for low multipoles (large angular scales) and translate to a modest enhancement of power, although small when compared to uncertainties coming from cosmic variance. 
Therefore, anisotropies do not alter significantly the best-fit value of the six free parameters of the  standard (Lambda cold dark matter) model. We have checked this by running a Markov chain Monte Carlo analysis  \cite{lewis}, using $TT$, $EE$, and $TE$ data \cite{planckdata}.  In contrast, correlation functions for $\ell\neq \ell'$ are a smoking gun for anisotropies \cite{peloso}. In Fig.~\ref{fig:DTT} we also show one of them, namely $C_{\ell \ell+2, 00}^{TT}$,  as an illustrative example. Other values of $\ell,\ell',m,m'$ produce  similar results. Our result for $C_{\ell \ell+2, 00}^{TT}$ is in agreement with the quadrupolar modulation observed by {\it Planck}.

\begin{figure}[h]
{\centering     
\includegraphics[width = 0.5 \textwidth]{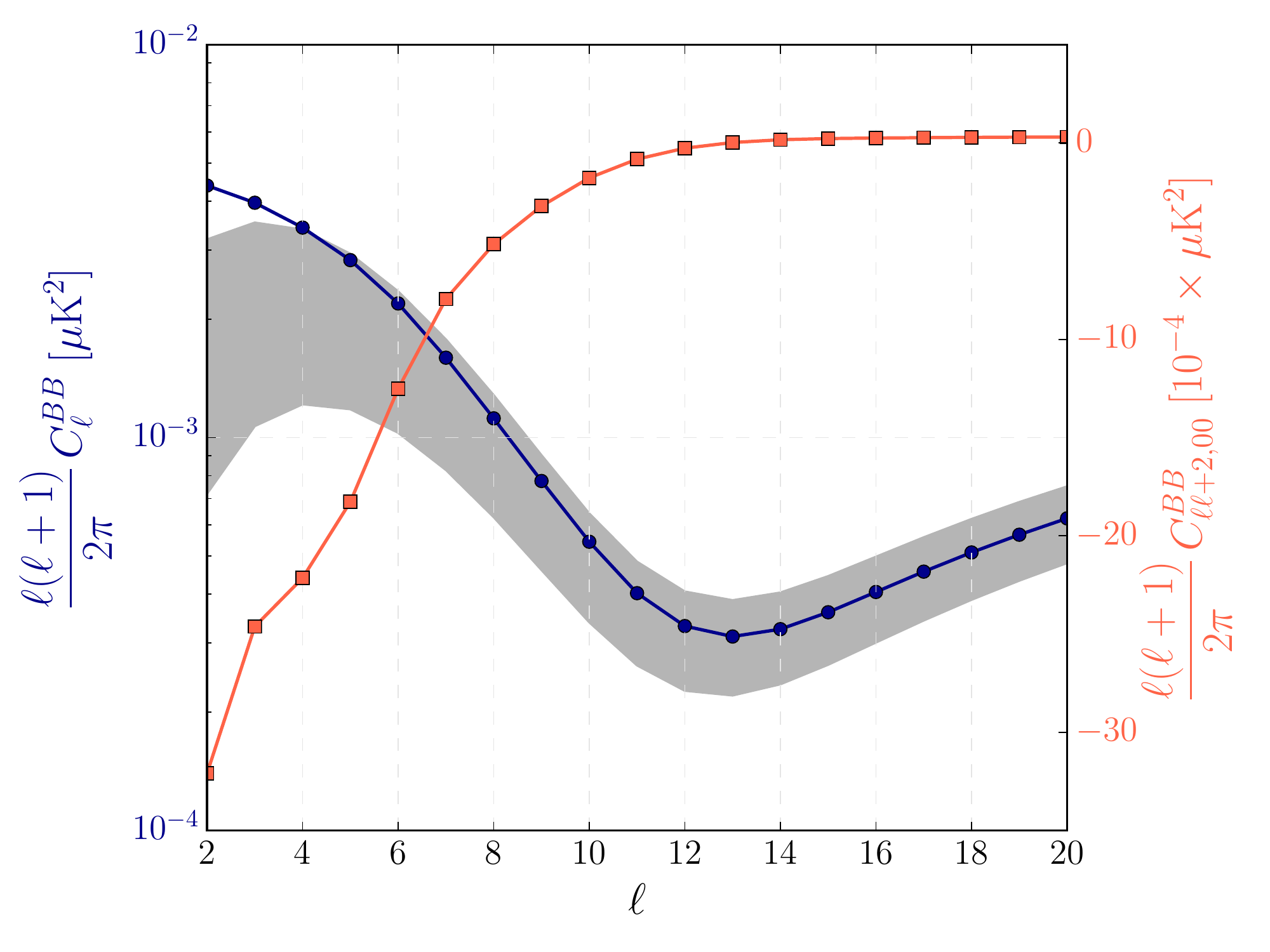} 
}
\caption{Left axis: $B$-$B$ polarization angular correlation function $C_{\ell}^{BB}$ (dotted blue line), and the predictions from isotropic inflation with cosmic variance (shaded region), for comparison. Right axis: Off-diagonal component of the $B$-$B$ polarization correlation function $C_{\ell \ell+2, 00}^{BB}$ (red line with squares).}
\label{fig:DBB}
\end{figure}

(ii) $E$-$E$, $B$-$B$, and $T$-$E$ correlations. The conclusions are similar to  the $T$-$T$ case. Namely, these correlations are different from zero only for $\ell+\ell'$ even, and the main departures from the isotropic model appear for low multipoles and for $\ell\neq \ell'$. As an example, we plot in Fig.~\ref{fig:DBB}  $C_{\ell}^{BB}\equiv \frac{1}{2\ell+1} \sum_{m=-\ell}^\ell (-1)^mC_{\ell \ell, m -m}^{BB}$ and $C_{\ell \ell+2, 00}^{BB}$. The latter has an important contribution from the entanglement between tensor perturbations with different polarizations. 

\begin{figure}[h]
{\centering     
  \includegraphics[width = 0.5 \textwidth]{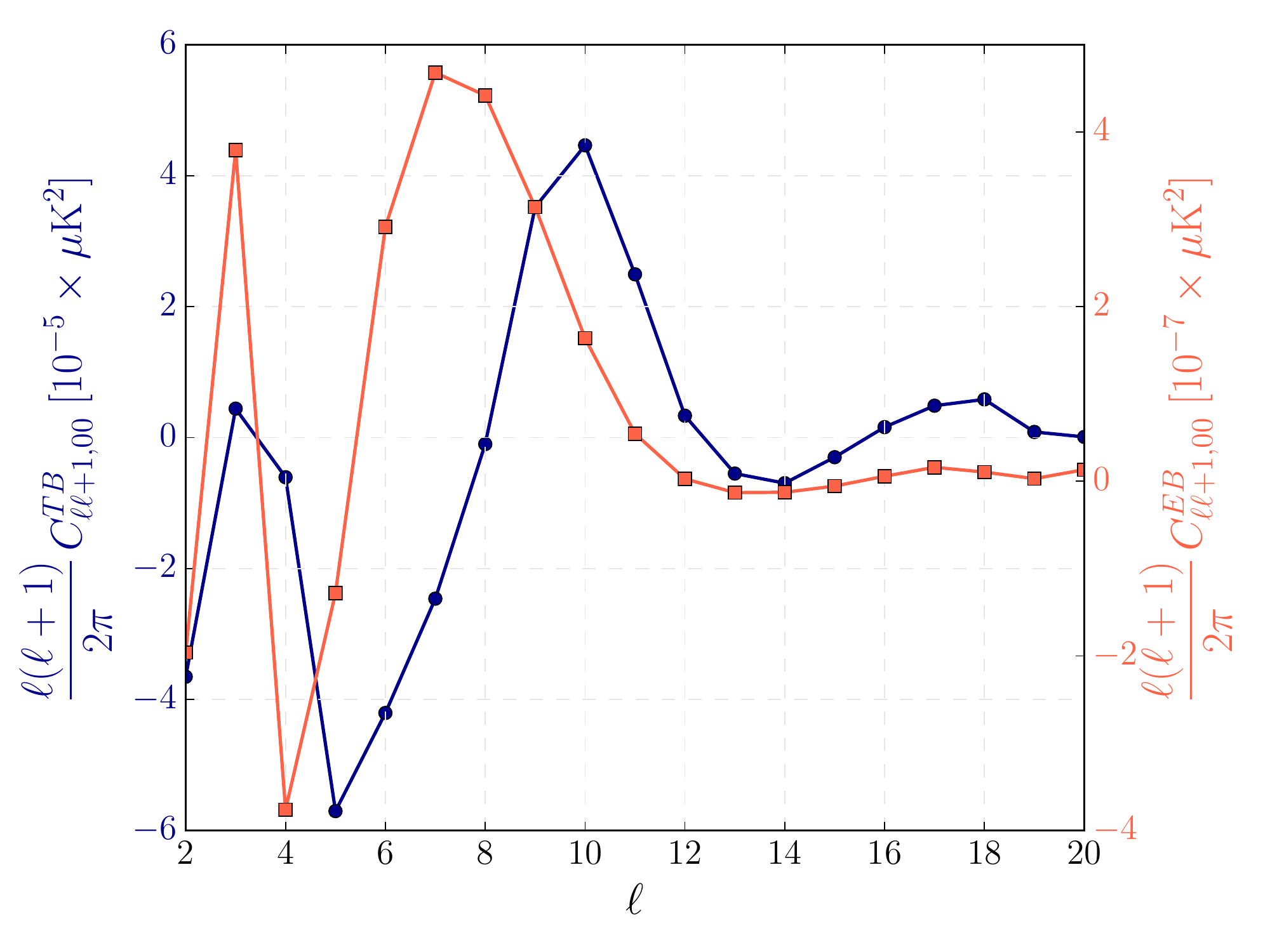}  
}
\caption{$T$-$B$  (left axis, dotted blue line), and $E$-$B$ (right axis, red line with squares) correlation functions for $\ell'=\ell+1$, and $m=0=m'$. The isotropic counterpart  is identically zero. } \label{fig:TB}
\end{figure}

(iii) $T$-$B$ and $E$-$B$. Because the $B$-polarization field is a pseudoscalar, while $T$ and $E$ are parity even,  these correlations vanish in a parity invariant theory unless $\ell+\ell'$ is odd. Since isotropy would also imply $\ell=\ell'$, all these correlations vanish  in the standard cosmological model.  Fig.~\ref{fig:TB}  shows $C_{\ell \ell+1, 0 0}^{T, B}$ and $C_{\ell \ell+1, 0 0}^{E, B}$ in our model. They originate  {\it exclusively} from the entanglement between scalar and the two tensor  modes. 

In the standard theory of inflation  the  amplitude of  tensor perturbations depends on the choice of $V(\phi)$. This freedom remains in our model. We have chosen the parameters in $V(\phi)$ that best fits existing data, but a different choice of $V(\phi)$ would change the amplitude of $B$-modes. Our invariant prediction for them is, therefore, the  magnitude of anisotropies relative to their overall amplitude.

The computational difficulty of these calculations comes from the need to  resolve the angular dependence of the primordial power spectra $\mathcal{P}_{s,s'}(\vec k)$ or, equivalently, to decompose $\mathcal{P}_{s,s'}(\vec k)$ in spherical harmonics with spin weight $s-s'$. This is a demanding task---the calculation of these plots takes about a week on a 96-core high performance computer (we use the numerical library \cite{num-lib}).

Our analysis  shows that the quadrupolar modulation of the $T$-$T$ spectrum observed by {\it Planck} \cite{plnck2018} could be a remnant from an anisotropic pre-inflationary phase, rather than a statistical fluke. Furthermore, we predict that this modulation comes together with concrete effects in the $E$-$E$, $T$-$E$, $B$-$B$, $T$-$B$ and $E$-$B$ correlation functions,  which provide a  way to test our ideas (further details omitted here can be found in \cite{aos2}).

{\bf Discussion.} 
The  merits of this Letter are as follows: (i) To introduce a Hamiltonian formulation of gauge invariant perturbations in Bianchi I spacetimes, and to implement the mathematical framework in a publicly available computational algorithm \cite{math-nb,num-lib,xart}. (ii) To formulate an exact quantization of the coupled system of linear  perturbations, and to use this formalism to compute the entanglement between scalar and tensor perturbations that anisotropies generate.  (iii) To embed this  theory within a quantization of the Bianchi I geometry, extending in this way  previous studies on quantum cosmology to anisotropic scenarios, a task that has remained elusive due to the complexity of the system. (iv) To show that perturbations can retain memory of the preinflationary universe, although the anisotropies in the background geometry quickly dilute during inflation. This memory is codified in the form of anisotropic correlation functions and quantum entanglement between the different types of perturbations.  (v) Finally, and most importantly, we have explained a possible origin for the nonzero quadrupolar modulation observed by {\it Planck}, and made concrete predictions for $E$-$E$, $B$-$B$, $T$-$E$, $T$-$B$ and $E$-$B$ correlations in the CMB. Although {\it Planck}'s observations of the $T$-$T$  quadrupole alone are not significant enough to declare the detection of anisotropic physics, a detailed search for the  effects we describe in the $E$-$E$, $T$-$E$ correlations (that {\it Planck} has already partially done), and particularly in $B$ polarization, could boost the  significance of the detection. Some of the values we predict, particularly the ones involving $T$-$B$ and $E$-$B$ correlations, are small and probably difficult to observe, but others are not, and could  be measured by the next generation of CMB polarization observatories, such as CORE \cite{core}. 
 
Furthermore, although we have worked within loop quantum cosmology, we expect our conclusions to be valid for other theories that predict a similar bounce (see, e.g., \cite{Mukhanov,6,Shtanov:2002mb}).\\

{\bf Acknowledgements.} 

\acknowledgments{We have benefited from  discussions with Abhay Ashtekar,  Mar Bastero-Gil, Brajesh Gupt, Guillermo A. Mena Marug\'an, Jorge Pullin, Parampreet Singh and Edward Wilson-Ewing. This work is supported by the NSF CAREER grant PHY-1552603, Project. No. FIS2017-86497-C2-2-P of MICINN from Spain and from the Hearne Institute for Theoretical Physics. V. S. was also supported by Louisiana State University and the Inter-University Centre for Astronomy and Astrophysics during different stages of this work. Portions of this research were conducted with high performance computing resources provided by Louisiana State University (http://www.hpc.lsu.edu).}

\end{document}